\documentclass[journal,letterpaper]{IEEEtran}
\usepackage[dvips]{graphicx}
\usepackage{url}
\newlength{\picwidth}
\begin{document}
%
\title{Demonstration of fine pitch FCOB (Flip Chip on Board) assembly based on solder bumps at Fermilab}
\author{M.~Trimpl, E.~Skup, R.~Yarema, J.C.~Yun\\\small{\textit{Fermi National Laboratory, Batavia, IL, 60510, USA}}}
\maketitle
\begin{abstract}
Bump bonding is a superior assembly alternative compared to conventional wire bond techniques.
It offers a highly reliable connection with greatly reduced parasitic properties.
The Flip Chip on Board (FCOB) procedure is an especially attractive packaging method for applications requiring a large number of connections at moderate pitch.
This paper reports on the successful demonstration of FCOB assembly based on solder bumps down to 250\,$\mu$m pitch using a SUESS MA8 flip chip bonder at Fermilab.
The assembly procedure will be described, microscopic cross sections of the connections are shown, and first measurements on the contact resistance are presented.
\end{abstract}


\section{Introduction}
\IEEEPARstart{F}{lip} chip on board is an advanced packaging method where the bare chip die is directly mounted onto a PCB using bump bond connections.

Generally, bump bonding offers a much higher connection density per chip area compared to wire bond approach where the bond pads are typically arranged around the chip edge. 
Since the bump bonds are intrinsically kept very short, parasitics in terms of resistance, capacitance and inductance are reduced to a minimum.
Low inductance connections are particularly critical for high speed applications.
As an example, a typical wire bond connection shows an inductance of about 1\,nH per mm \cite{CITE_WIREBONDING}, whereas a bump bond could provide 50\,pH only.
The reduction of resistance and capacitance is in the same order of magnitude.
\begin{figure}[h]
  \centering
  \includegraphics[width=\picwidth]{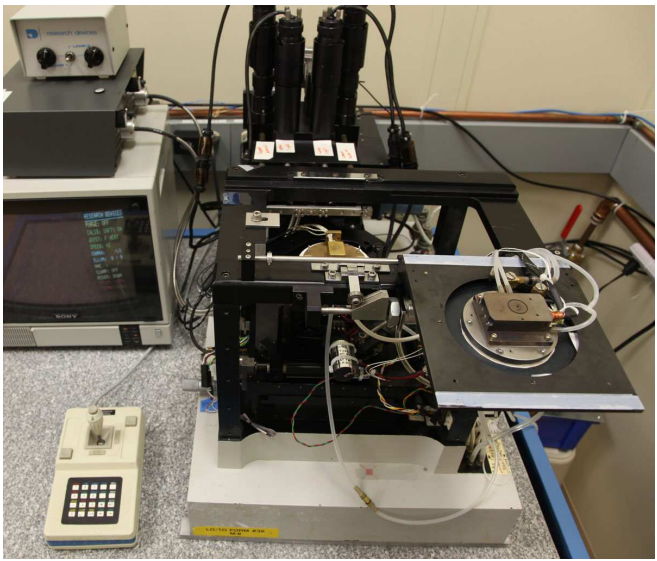}
  \caption{Picture of the Suess MA8 flip chip bonder setup at Fermilab.}
  \label{Suess_FChipper}
\end{figure}

For specific applications where high current handling capability is required, bump bonding presents crucial advantages most of all. 
It may be feasible to conduct a high current through a wire bond\footnote{A single 10\,mm long, 1\,mil.~Al/Si wire bond can carry up to 500mA.} to the chip edge, however, the challenge still remains to distribute this current efficiently to the CMOS circuitry on the chip level.
Bump bonding intrinsically solves this topological challenge as the current can be delivered equably to the whole chip area.
High current applications will also favor solder bump connection as they reflect the most robust and reliable bonding providing lowest resistance.
Additionally, the whole backside of the die can be used to dissipate heat generated by the circuitry on the chip.
In industry, one of the most prominent examples exploiting these features are power hungry microprocessors.
For the SPi chip, which has been designed at Fermilab and will be outlined later, these were also the most important reasons to use bump bonding assembly.

To generally explore vertical assembly methods at Fermilab, a SUESS MA8 flip chip bonder has been purchased.
This paper will focus on recent achievements using solder bump techniques.
Besides metallic bumps that form a solder connection, anisotropic materials as well as isotropic conductive polymer technologies, such as stencil printing are being explored by industry as well and are mentioned for completeness.

The setup that has been used for the assemblies to be described is shown in Figure~\ref{Suess_FChipper}.
\section{Bonding test and bump resistance}
\begin{figure}
	\centering
  \includegraphics[width=\picwidth]{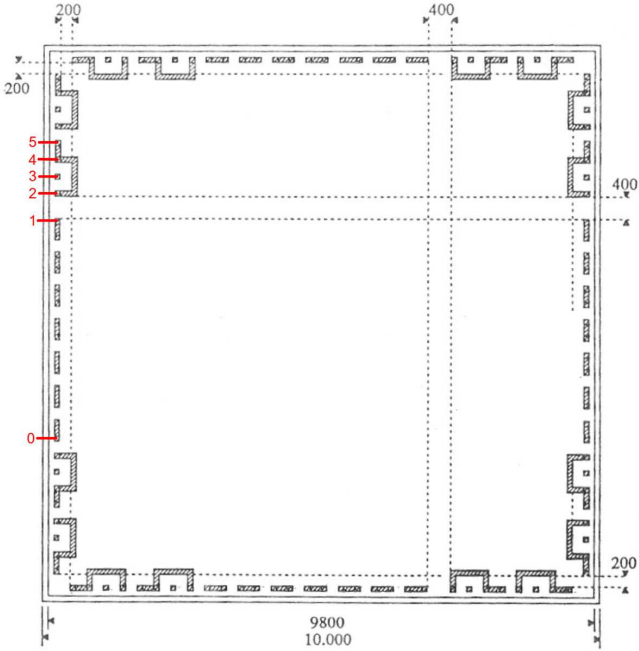}
  \caption{Layout of the various structures on the test chip. Sketch courtesy of IZM Berlin. The regular bump pitch is 300\,$\mu$m unless noted otherwise. Dimensions in $\mu$m.}
  \label{IZM_overview}
\end{figure}

For initial bonding attempts, dedicated test chips were purchased from IZM Berlin \cite{CITE_IZMBerlin}.
These test chips are $\mathrm{10\times10\,mm^2}$ large and contain different bump structures along the chip edge.
The bumps are SnAgCu eutectic with approximately 150\,$\mu$m diameter and use a 5\,$\mu$m Ni/Au under bump metalization (UBM).
Figure~\ref{IZM_overview} shows the floor plan of the chip.
In each corner, there are 4 hook-like structures composed of 4 bumps.
In between of these structures there are 14 bumps forming one part of a daisy chain on the chip level.
The structures along each edge are identical.
The regular bond pitch is 300\,$\mu$m.
A dedicated PCB has been designed to host the test chip.
It accesses all points of the hook-like structures (indicated by points (2) through (5) in Figure~\ref{IZM_overview}).
It also completes the daisy chain (accessible via points (0) and (1) as shown in Figure~\ref{IZM_overview}) on PCB level.

\begin{figure}[b]
	\centering
  \includegraphics[width=\picwidth]{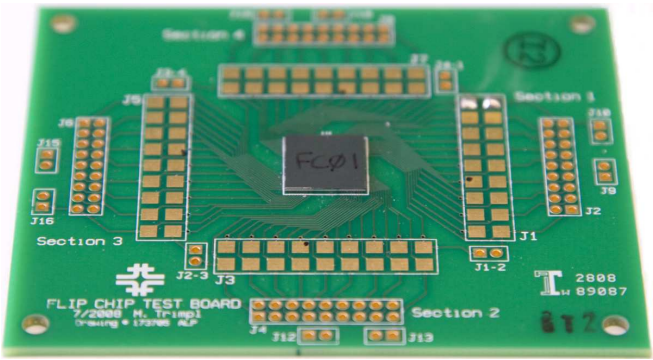}
  \caption{Flip chip test board using standard FR4 material populated with $\mathrm{10\times10\,mm^2}$ chip from IZM Berlin.}
  \label{IZM_FCboard}
\end{figure}

Figure~\ref{IZM_FCboard} shows a picture of the test chip successfully mounted onto the PCB.
For the assembly the following procedure has been used:
\begin{itemize}
  \item{flip chip at a temperature of 110\,$^\circ$C and apply an equivalent pressure of 1500\,g onto the chip}
  \item{tack the chip by raising the temperature to 180\,$^\circ$C and holding it for 1\,min}
  \item{release pressure of upper chuck}
  \item{reflow at 210\,$^\circ$C for 3\,min}
\end{itemize} 
After assembly, connectivity of all bump connections was observed for the full chip.
However, a connectivity test does not show shorts between neighboring bumps.
If the bumps get squished too much so that they touch each other, connectivity of single bumps as well as of the daisy chain would still be maintained.
To detect shorts between bumps, the hook-like structures can be used by confirming open-circuits between points (3-2) and (3-4).
For all structures on the chip, no shorts were observed.

The setup was also used to determine the bump resistance.
The resistance of the full daisy chain has been measured to be $\rm{R_{daisy}=1061\pm1\,m\Omega}$.
This value reflects:
\begin{displaymath}
  \mathrm{R_{daisy}=R_{par01}+14\,R_{bump}+6\,R_{PCB\_line}+7\,R_{chipline},}
\end{displaymath}
with
\begin{itemize}
  \item{$\mathrm{R_{par01}}$ being the overall parasitic resistance going to the bump bond pads (0) and (1) of the daisy chain (see Figure~\ref{IZM_overview}),} 
  \item{$\mathrm{R_{PCB\_line}}$ and $\mathrm{R_{chipline}}$ being the resistances of the traces that connect the bump pads within the daisy chain on the PCB board and on the chip, respectively,}
  \item{ and finally, 14 bump bond connections $\mathrm{R_{bump}}$ within the daisy chain.}
\end{itemize}
$\mathrm{R_{par\_01}}$ has been measured separately on a bare PCB to be $\mathrm{129\pm1.4\,m\Omega}$.
$\mathrm{R_{chipline}}$ can be determined by subtracting the measured resistance between the points (2-5) and (2-4) of the hook-like structures.
This yields $\mathrm{R_{chipline}=124.5\pm1.3\,m\Omega}$.
The PCB traces are 0.3\,mm long and 0.1\,mm wide and use 1\,oz.~copper\footnote{0.476\,m$\Omega$ per square for 1\,oz.~copper}. 
Their resistance can be estimated to merely contribute $\mathrm{R_{chipline}=1.43\,m\Omega}$.  
This results in a resistance of a single bump connection of
\begin{equation}
  \mathrm{R_{bump}=3.7\pm0.8\,m\Omega.}
\end{equation}
For comparison, a rather short wire bond (2\,mm long using 1\,mil.~diameter wire) typically shows $\mathrm{\sim100\,m\Omega}$ resistance \cite{CITE_WIREBONDING}.
This difference is evidently due to the geometrical superiority of the assembly rather than based on any material property. 
\section{Microscopic inspection of the bump bond assembly}
\begin{figure*}
  \centering
  \includegraphics[width=179mm]{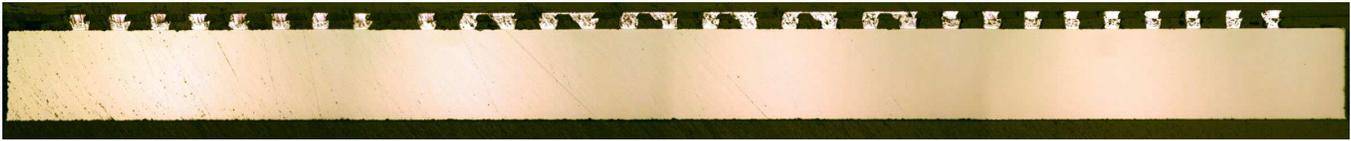}
  \caption{Microscopic image showing a cross section of one side (length 10\,mm) of the test PCB. The bottom layer is the chip die with the PCB situated on top. The eight single bumps on each side belonging to the two hook-like structures are shown, as well as the 14~bumps of the daisy chain in the middle.}
  \label{IZM_Cross}
\end{figure*}

\begin{figure}
	\centering
  \includegraphics[width=\columnwidth]{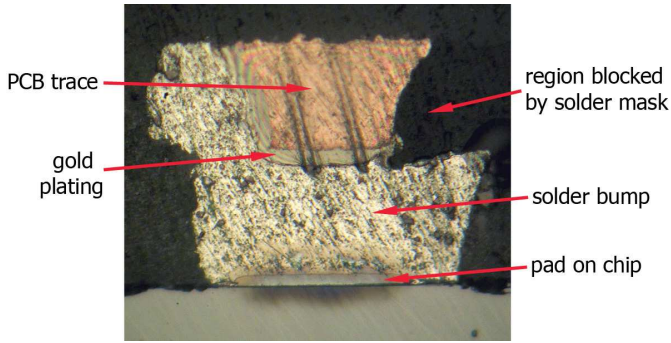}
  \caption{Close-up view of the 4th single bump from left in Figure~\ref{IZM_Cross}. The region where the solder mask prevented gold plating and bump attachment is clearly visible. Solder bump width is about 150\,$\mu$m.}
  \label{IZM_singlebump}
\end{figure}

\begin{figure}
  \centering
  \includegraphics[width=\picwidth]{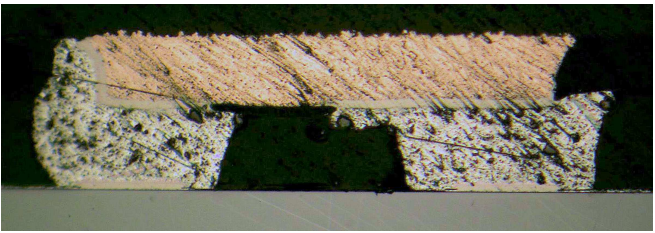}
  \caption{A separate microscopic image showing a single element of the daisy chain. The bump pitch is 300\,$\mu$m}
  \label{IZM_daisychain}
\end{figure}

Microscopic images of the assembly have been taken to verify the alignment and bumping quality.
The chips were potted using cold mount epoxy technique and the stack cut to obtain cross sections \cite{CITE_EAG}.
Figure~\ref{IZM_Cross} shows the full cross section along one chip edge. 
The 8 single bumps on each side belonging to the two hook-like structures can be identified, as well as the 14~bumps of the daisy chain in the middle.
In the picture the actual chip die is situated at the bottom and the PCB is shown on top.
Figure~\ref{IZM_singlebump} illustrates a close-up view of a single bump of the hook-like structures.
Figure~\ref{IZM_daisychain} shows a detailed view of one element of the daisy chain.
The cut for this particular cross section has been made a little deeper than that shown in Figure~\ref{IZM_Cross}, so that the full thickness of the PCB trace is shown.

A few conclusions can be drawn from the microscopic inspection.
First of all, the position alignment of the flip chipping itself is very good, much better than required by a 300\,$\mu$m bump pitch.
On the other hand, a closer look at Figures~\ref{IZM_singlebump} and \ref{IZM_daisychain} shows that the gold plating on the PCB is not covering the copper trace completely, meaning not on both sides of the pad.
This is due to mask alignment limitation of the PCB fabrication and has been confirmed by inspection of an unbumped PCB under the microscope.
Consequently, during the reflow step the bump ball attaches to the gold plating as it is on the PCB (the solder mask blocked the deposition of the gold plating as well as any connection of the bump to the copper trace). 
This leads to a solder connection on one side of the pad only and a slightly uneven appearance of the bump.
However, for typical bump pitches used in FCOB (200-300\,$\mu$m) this matter will not limit the assembly. 
For finer pitches it can be concluded that standard FR4 substrate is probably not suitable and manufacturing processes with higher accuracy have to be considered.
However, it is questionable if a finer pitch is reasonable for mounting a chip on a PCB as traces of equivalent pitch have to be routed on the board as well.
Also reliability issues due to mechanical stress on the sandwich may become more relevant.
\section{Assembly of the SPi~001 chip}
\begin{figure}[!b]
	\centering
  \includegraphics[width=\picwidth]{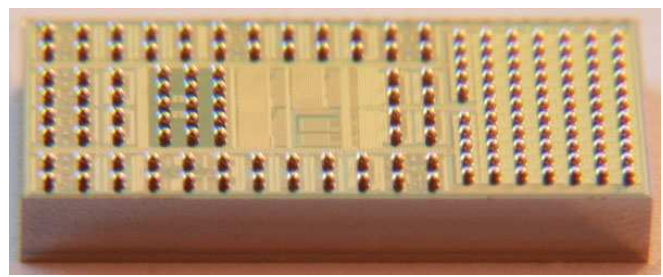}
  \caption{Microphotography of the $\rm{5.7\times2.8\,mm^2}$ large SPi~001 chip with SnPb bumps.}
  \label{SPi_bumps}
\end{figure}

The next step for FCOB assembly was to use the SPi (Serial Powering Interface) chip \cite{CITE_SPiTrimpl} as a real application.
The SPi has been explicitly designed to take advantage of bump bond features. 
The main motivation was the ability to handle currents up to 4\,A while providing a very low impedance path.
The bare chip is shown in Figure~\ref{SPi_bumps}. 
The bumps have appropriate UBM and have been placed by TSMC foundry.
Eutectic bumps composed of 37\,Pb~/~63\,Sn with $\sim100\,\mu$m diameter were chosen.
The bump pitch is not equal over the chip.
A 325\,$\mu$m pitch is used in the left hand pad frame and a 250\,$\mu$m pitch is used in the bump array on the right side.

\begin{figure}
	\centering
  \includegraphics[width=\picwidth]{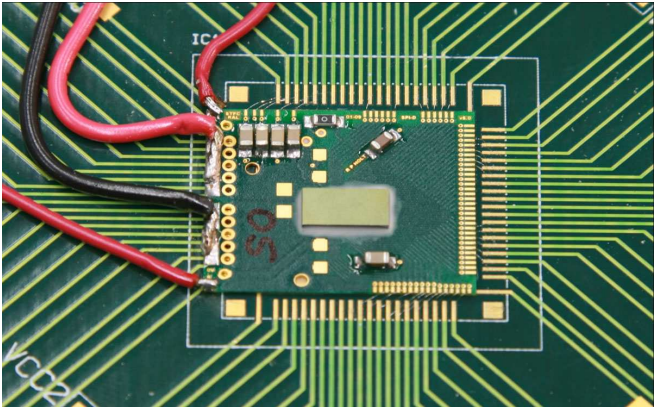}
  \caption{Photography of the $\rm{15\times18.5\,mm^2}$ daughter board with mounted SPi. The board is glued and wire bonded to a test PCB.}
  \label{SPi_daughter}
\end{figure}

A dedicated daughter board \cite{CITE_DaughterRAL} has been fabricated to host the SPi chip and some additional passive components.
Figure~\ref{SPi_daughter} shows the SPi~001 chip successfully bumped onto the daughter board.
In order to communicate with the chip signals are wire bonded to the test PCB to which the daughter board is glued to.
For all high current paths cables are directly soldered to the daughter board.

A slightly different procedure has been used to mount the SPi chip:
\begin{itemize}
  \item{flip chip at a temperature of 100\,$^\circ$C and apply an equivalent pressure of 500\,g}
  \item{tack the chip by raising the temperature to 150\,$^\circ$C and holding it for 3\,min}
  \item{release pressure of upper chuck}
  \item{reflow at 195\,$^\circ$C for 1.5\,min}
\end{itemize} 
The most important difference compared to the assembly of the IZM chip is the reflow temperature. 
In both cases it has been determined initially by raising the temperature while applying little pressure to the chip and observing when the stack collapses.
Note that the difference in equivalent pressure for tacking the chip is due to the different chip area.
  
Epo-Tek T7110 \cite{CITE_Epoxy} has been used to encapsulate the chip. 
The epoxy is designed to act as an underfill material and capillary flows under the chip between the solder bumps.
After application it has been cured for 2\,hours at 80\,$^\circ$C.  
Mounted chips have also been temperature cycled between -40 and +100\,$^\circ$C with no observed effect on connectivity.

One major component of the SPi~001 is a shunt regulator that basically acts as a constant voltage source.
The circuit has been extensively tested and a dynamic impedance of less than 50\,m$\Omega$ measured for a wide frequency range (up to 10\,MHz).
Based on circuit and parasitic simulations, about half of this impedance is due to the regulation circuit itself and about half is due to parasitic effects. 
This demonstrates the low resistance and also low inductance that is offered by the flip chip assembly.
Moreover, a current handling capability of 4\,A of the chip has been demonstrated.
\section{Comment on thermal conductivity}
One of several advantages of bump bonding the SPi chip to the PCB is that the backside is fully accessible for cooling purposes.
This section will briefly illustrate thermal considerations for bump bonded devices using the SPi chip as an example. 

Depending on the operational circumstances of the SPi, currents on the order of 1\,A or more at the nominal voltage of 2.5\,V may occur.
Since CMOS circuitry is not implanted deeply into the silicon substrate, this power (2.5\,W) will be generated very close to the front side of the chip and, hence, has to be dissipated efficiently.
Fortunately, silicon has a relatively good thermal conductivity. 
It can be computed \cite{CITE_SIthermal} to be $\mathrm{\lambda_{Si}}$=128.5\,W/mK at 350\,K (for comparison copper is about 380\,W/mK).
For a given geometry, the thermal conductance $\sigma$ of an object can be calculated according to: 
\begin{equation}
  \rm{\sigma = \frac{P}{\Delta T} = \lambda \cdot \frac{A}{d}}
\end{equation}
with A representing the dissipating surface area and d being the thickness of the object.
Using $\rm{A=5.7\times2.8\,mm^2}$ as chip area and d=720\,$\mu$m for the thickness yields 
\begin{displaymath}
  \mathrm{\sigma=2.85\,W/K}
\end{displaymath}
for the SPi~001 chip.
That means that for the discussed power consumption of 2.5\,W, the temperature difference between the circuitry at the front side of the chip and the back side would be less than 1\,K.
Furthermore, efficient cooling measures can be arranged to dissipate the power further on.
This could be cooling gas gently flowing over the die or a more aggressive cooling element physically attached to the chip.

In addition to that, the bump balls (with or without underfill) will provide cooling via the front side.
\section{Conclusion and Outlook}
In-house flip chip on board assembly has been demonstrated at Fermilab down to a bonding pitch of 250\,$\mu$m.
An epoxy underfill has been used to enhance mechanical stability.
Temperature cycles between $-40\,^\circ$C and $+100\,^\circ$C have been conducted with no noticeable effect on connectivity.   

Based on microscopic studies it is reasonable to say that FCOB assemblies down to 200\,$\mu$m are possible using standard FR4 PCB material in the setup.
With the alignment precision of the equipment being much better, the current limiting factor in bonding pitch is the mask alignment of the PCB fabrication.
More accurate fabrication processes, such as those used for ceramic boards, would enable even finer pitches. 

The resistance of a single bump connection has been measured to be below 4\,m$\Omega$ which is considerably lower than what can be achieved using wire bond technology.
Although not being explicitly measured, inductance and capacitance of the connection is also significantly lower for the bump bond assembly. 

An alternative assembly using anisotropic adhesive tape\footnote{This is not being considered for the SPi chip but may be beneficial for other applications.} instead of solder bumps is under investigation.

\section*{Acknowledgment}
The authors would like to thank Angela Prosapio for designing the test PCB and Albert Dyer for researching and applying the underfill epoxy.

\end{document}